\documentclass[%
 aip,
 amsmath,amssymb,
 reprint,%
]{revtex4-1}

\usepackage{graphicx}% Include figure files
\usepackage{dcolumn}% Align table columns on decimal point
\usepackage{bm}% bold math
\usepackage[utf8]{inputenc}
\usepackage[T1]{fontenc}
\usepackage{mathptmx}

\begin{document}

\preprint{AIP/123-QED}

\title{Effective transport properties of conformal Voronoi-bounded columns via recurrent boundary element expansions}
% Force line breaks with \\

\author{Matthew D. Arnold}
\email{Matthew.Arnold-1@uts.edu.au}
\affiliation{%
 School of Mathematical and Physical Sciences, University of Technology Sydney.
}%

\date{\today}% It is always \today, today,

\begin{abstract}
Effective transport properties of heterogeneous structures are predicted by geometric microstructural parameters, but these can be difficult to calculate.  Here, a boundary element code with a recurrent series method accurately and efficiently determines the high order parameters of polygonal and conformal prisms in regular two-dimensional lattices and Voronoi tessellations (VT).  This reveals that proximity to simpler estimates is associated with: centroidal VT (\textit{cf} random VT), compactness, and VT structures (\textit{cf} similarly compact semi-regular lattices).  An error in previously reported values for triangular lattices is noted.
\end{abstract}

\maketitle

\section{\label{sec:Intro}Introduction}

\begin{figure}[h]
\includegraphics[width=\columnwidth]{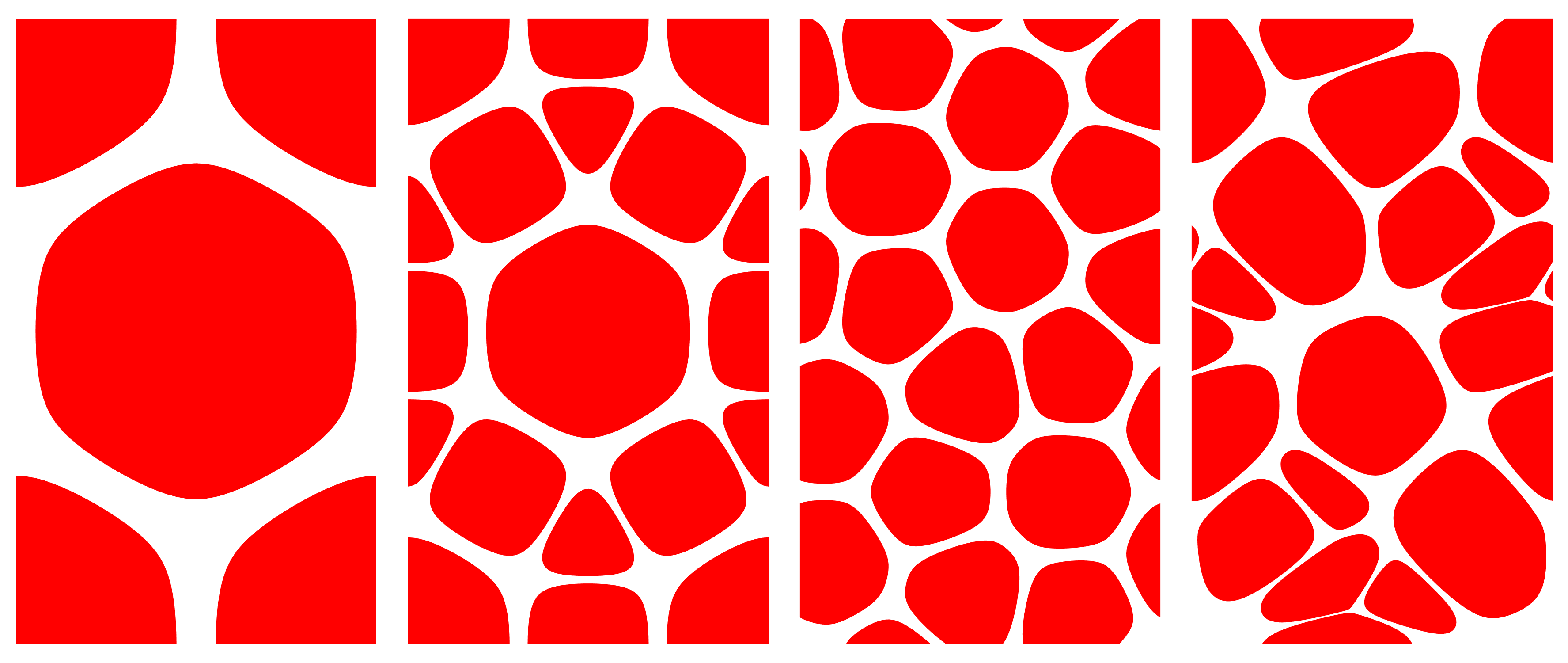}
\caption{\label{fig:Structures} Examples of conformal inclusions in various lattices (at \textit{f}=0.75): left to right: hexagonal (regular), 3.4.6.4 (semi-regular), CVT, and random VT (evolving from centroids).}
\end{figure}

The effective transport model for random heterogeneous media is a topic of ongoing interest \cite{Torquato2002,Milton2002}, because it is valuable for understanding the limits of application of these structures.  Most earlier studies focused on regular structures since investigation of more realistic psuedo-random structures can be computationally expensive.  Even so, there have been some useful early approaches to this problem including level-cut random fields\cite{Roberts1995} and random Voronoi tessellations\cite{Winterfeld1981percolation,Jerauld1984percolation}. Recently there has been a resurgence in interest in hyper-uniform disordered (HUD) structures\cite{TorquatoStillinger2003} that have ordering intermediate between crystals and random glasses, offering a number of useful properties such as enhanced isotropy\cite{Man2013isotropic}, distributed absorption\cite{Liu2018absorption} and wide bandgaps with dense bands\cite{Florescu2009designer}.  Centroidal Voronoi tessellations (CVT)\cite{Du1999Centroidal} are an important example of this class of geometry\cite{Klatt2019}, and are a plausible model of some real structures such as anodic alumina pores\cite{Previdi2019}.  Most studies of effective transport in random VT have focussed on inclusions that lie near the extremes, i.e. either strut-like\cite{TorquatoGibiansky1998} or disc-like \cite{Vrettos1989effective} with the response of the latter complicated by penetrability of the discs even at low fill factor\cite{TorquatoLado1992}.   In general, transport at all fill-factors can be unambiguously investigated using non-percolating contours\cite{Arnold2017}, including some that achieve the optimal bounds\cite{Vigdergauz1994two, Vigdergauz1996rhombic, Vigdergauz1999energy, hyun2002optimal}.  Since the Vigdergauz structures by definition have trivial microstructural factors and are generally difficult to generate, this article is restricted to less optimal conformal contours \cite{Driscoll1996algorithm}, and polygonal contours, in VT.   Examples of the geometries with conformal inclusions are shown in Fig.~\ref{fig:Structures}: while the primary focus here is statistical transport of random VT, a selection of regular and semi-regular lattices is surveyed.  To improve the accuracy of the estimated transport of (piecewise) smooth shapes, a boundary element method is used to recurrently determine series expansion coefficients in the form of microstructural factors.    This article discusses presents the recurrent formulation of boundary elements, discusses literature on transport in VT and generation of CVT, and finally analyses third order parameters for inclusions in semi-regular lattices and random VT.

\section{\label{sec:Theory}Theory}
\begin{figure}[h]
\includegraphics[width=0.8\columnwidth]{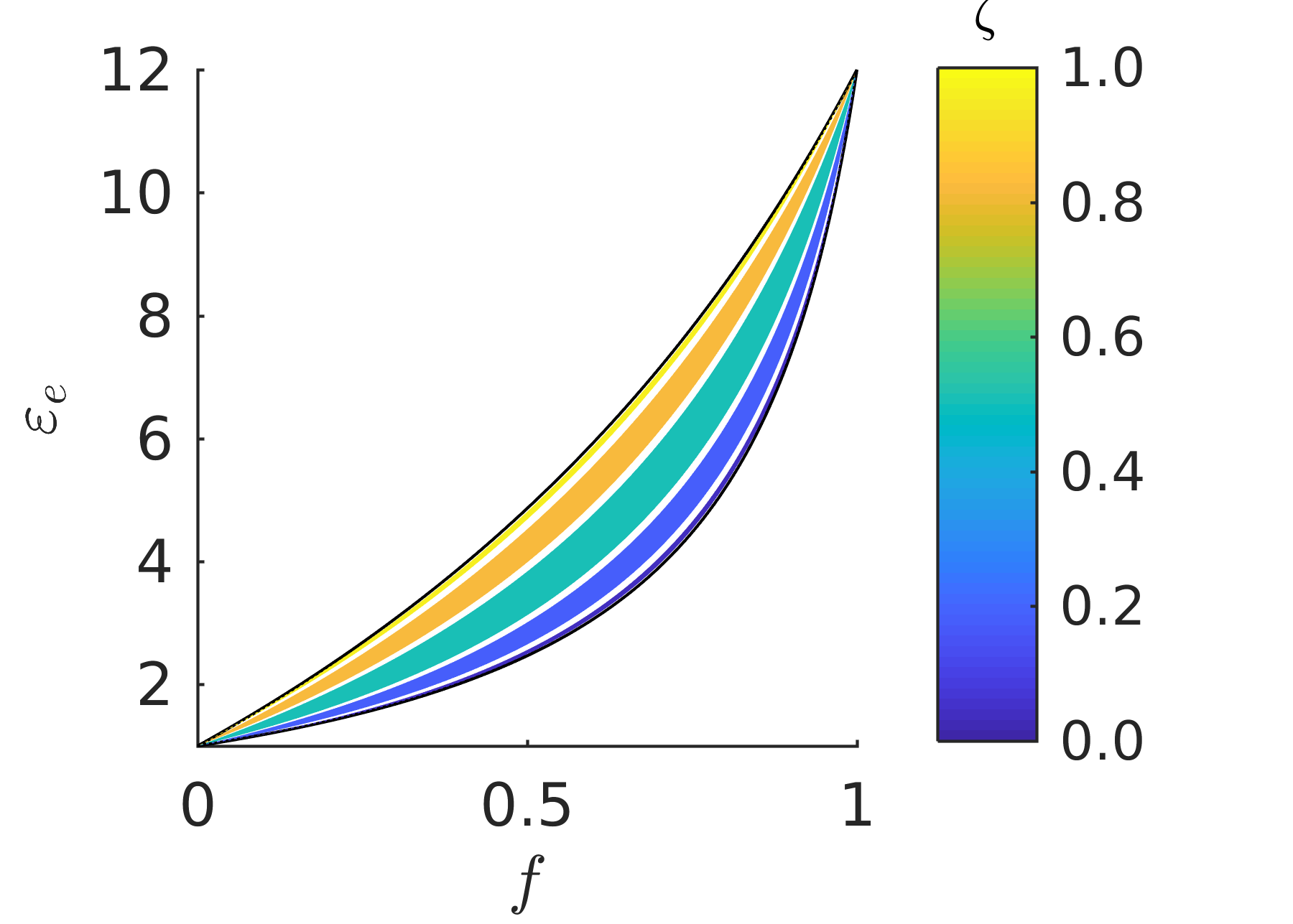} \caption{\label{fig:Bounds} Demonstration of the 4\textsuperscript{th} order isotropic bounds on effective transport $\varepsilon_e$ for various values of the third-order microstructural parameter $\zeta$ (0.03, 0.18, 0.50, 0.82, 0.97), when the contrast ratio is $\varepsilon=12$.  These are in turn are bounded by the 2\textsuperscript{nd} order Hashin-Strikman bounds (black lines), which only require knowledge of $f$.  For reference, $\zeta < 0.2$ for triangles and  $\zeta < 0.02$ for hexagons in their natural lattices.}
\end{figure}

The primary focus in this article is the effective conductivity (thermal or electrical or permittivity or permeability).  Separating the geometric and material contributions has two popular approaches.  Spectral approaches \cite{Bergman1979} are useful for understanding plasmon resonances\cite{Arnold2017}, but are more difficult to apply near singular points (such as percolation or sharp points), as discrete resonances merge into distributions\cite{Djordjevic1996spectral}. Alternatively, in the dielectric regime it is more appropriate to employ a series approach \cite{Brown1955, Sen1989, Engstrom2005}, and near sharp points the convergence of the series parameters seems to be more reliable than that of the spectral modes.   In this approach, which is used in this article, a sequence of so-called microstructural parameters determine the response relative to optimal bounds.  It is known that even orders are trivial for isotropic structures, and the first (odd) order is fill-factor $f$, but higher order factors are more difficult to determine \cite{Sen1989}.  The third order parameter $\zeta$ is often the most important subject of calculation, since this primarily determines the proximity to the well-known Hashin-Strikman bounds (see Fig.~\ref{fig:Bounds}), but if percolation or sharp corners occur then higher orders also become important at high contrast. Recently we showed how to efficiently calculate these factors to arbitrary order \cite{Gali2019}, but used a structured grid that limited practical results to about fourth or fifth order.  Here, a recurrent approach with an implementation of a boundary element technique (BEM)\cite{Arnold2017} is developed with higher accuracy for piecewise smooth shapes.

The microstructural approach is based on expanding the relative effective permittivity as a power series in the relative bulk susceptibility
\begin{equation}
\varepsilon_e = 1 + \chi_e = \sum_{m=0}^{\infty} a_m \chi^m.
\label{eq:powerSeries}
\end{equation}
The direct series is not guaranteed to converge under all circumstances, however its Pade approximants are more reliable and analysis of these has produced important insights \cite{Gali2019}.  Once the coefficients $a_m$ in eq (\ref{eq:powerSeries}) are determined (and those for the inverse permittivity), we can use methods outlined previously \cite{Gali2019} to determine a sequence of bounds on the permittivity \cite{Sen1989}, then the bounds on the coefficients themselves \cite{Engstrom2005}, and finally the sequence of microstructural parameters in the range (0,1) that predict the effective permittivity relative to the previous bounds\cite{Gali2019}.

In the BEM, it is natural to start with the relative effective susceptibility
\begin{equation}
\chi_e = P \sigma / A,
\label{eq:collapseEq}
\end{equation}
where $A$ is the unit cell area, and $P$ is a dipole operator (e.g. $P_x=\int dr x$ ) acting on the surface charge density $\sigma$, given by
\begin{equation}
\sigma = [1/\alpha - G]^{-1} E_n,
\label{eq:directSol}
\end{equation}
consistent with the outwards-normal projection of the applied field $E_n$, surface polarizability $\alpha$ and interaction $G$.  Here we may choose surface polarizability
\begin{equation}
\alpha = [1/2+1/\chi]^{-1}.
\label{eq:polEq}
\end{equation}
The corresponding 2D interaction operator is
\begin{equation}
G = \frac{1}{2\pi} \int \vec {dr^\prime} \frac{\hat n \cdot (\vec r - \vec r ^\prime)}{| \vec r - \vec r ^\prime | ^ 2}
\label{eq:rawGreens}
\end{equation}
where $\vec r$ is a point on the surface and $\hat n$ the corresponding normal. Numerically implementing eq \ref{eq:rawGreens} requires careful treatment of self-singularity (e.g. via closure \cite{Mayergoyz2005}), lattice summation (e.g. in 2D, the Weierstrass $\zeta$ elliptic function removing extraneous terms\cite{Arnold2017} ), and consideration of the lattice termination.

To account for the sum termination we can consider the relationship between the applied field external to the termination ($E_0$) and the effective internal field $\langle E_e \rangle$ \cite{Jones1945Ellipsoid}:
\begin{equation}
E_0=(I+S_{bound}\chi_e)\langle E_e \rangle,
\label{eq:boundaryEq}
\end{equation}
where $S_{bound}$ is the depolarization of the termination, which can be conveniently chosen to correspond to a disc.  Comparing eq (\ref{eq:directSol}) and (\ref{eq:boundaryEq})
\begin{equation}
\chi_e^{-1} = E^{-1} [1/\alpha-\tilde G] P^{-1} A,
\label{eq:boundaryCorrected}
\end{equation}
where the termination-corrected lattice sum is
\begin{equation}
\tilde G = G - E S_{bound} P / A.
\label{eq:boundaryCorrection}
\end{equation}

Now, we can use successive substitutions \cite{Brown1955} of the self-consistent equation
\begin{equation}
\sigma = \alpha (E_n + \tilde G \sigma),
\label{eq:selfEq}
\end{equation}
into itself, yielding a Neumann series
\begin{equation}
\sigma = \sum_{m=0}^{\infty} \alpha [\tilde G \alpha]^m E_,
\label{eq:NeumannSeries}
\end{equation}
Assuming a binary composite, and writing recurrent geometric coefficients
\begin{equation}
q_m = P \tilde G^m E_n / A
\label{eq:auxCoeffs}
\end{equation}
we find the trivial coefficients $a_0=1$ and $a_1=f$ , and also generally (for $m \geq 1$)
\begin{equation}
a_m = \sum_{k=0}^{m-1} q_k (-1/2)^{m-1-k} \binom{m-1}{k}.
\label{eq:powerCoeffs}
\end{equation}
At this point the bounds and microstructural factors can be determined with techniques described elsewhere \cite{Gali2019}. In two dimensions, and if isotropy is assumed (which should hold on average for large enough VT), it can be shown that the third order parameter may be estimated with:
\begin{equation}
\zeta = \frac{4 a_3}{f(1-f)}-(1-f)
\label{eq:zetaIso}
\end{equation}

This parameter, which must lie in the range $0 \leq \zeta \leq 1$, leads to bounds \cite{Sen1989} which for isotropic two dimensional geometries are specified to 4\textsuperscript{th} order, simplifying to: 
\begin{equation}
\frac{\varepsilon_b}{\varepsilon_j}=\frac
{\varepsilon+\varepsilon_i(\varepsilon_{f}+\varepsilon_{1-\zeta})+\varepsilon_{f}\varepsilon_{\zeta}}
{\varepsilon+\varepsilon_j(\varepsilon_{1-f}+\varepsilon_{\zeta})+\varepsilon_{1-f}\varepsilon_{1-\zeta}}
\label{eq:fourthBound}
\end{equation}
where $\varepsilon_i$ and $\varepsilon_j$ take values $\varepsilon$ and 1 to be swapped to generate two bounds, and $\varepsilon_f=\varepsilon f + (1-f)$ and so on. Figure \ref{fig:Bounds} demonstrates that these bounds are very tight at extreme values of $f$ or $\zeta$, and it can be shown that the widest case ($f=1/(1+1/\surd \varepsilon)$ and $\zeta=1/2$) improves on the simpler 2nd order Hashin-Strikman bounds by a ratio $[(\surd \varepsilon - 1)/(\surd \varepsilon + 1)]^2$, and each subsequent odd parameter predicts the next even bound which improves by the same ratio.

In the results shown below, exact isotropy cannot be guaranteed, so the generalized procedure \cite{Gali2019} was used.  Note that care is required in interpreting results for rotated anisotropic structures.  For example, it is clear that low order bounds that do not incorporate autocorrelation are diagonal and by definition cannot bound the off-diagonal transport.  Further, the bounds on the second order coefficient $a_2$ are also diagonal and yet the tensor coefficient is not always diagonal.  Hence, in general microstructural tensors should be diagonalized to extract meaning from them.

\section{\label{sec:Literature}Transport in Voronoi Tessellations}
Voronoi tesselations are foam-like structures where each cell represents a region closest to “generator” points.  The effective transport of VT have been considered by various authors.  Although not directly VT, the third order parameter for identical circular inclusions in various distributions was compared\cite{TorquatoLado1992}, specifically fully penetrable discs with fully random positioning had higher $\zeta$ than hard discs which must necessarily be more ordered.  Both had significantly higher $\zeta$ than the structures presented in this article.  Thin strut VT (generated from hard-disc centers) were investigated in the context of elastic properties\cite{TorquatoGibiansky1998}, but they considered a single fill factor and did not determine either of the third order parameters ($\zeta$,$\eta$) required for elasticity.  Zhu Hobdell and Windel\cite{Zhu2001Disorder} studied the elastic properties of VT walls as a function of the effect of regularity (quantified by intergenerator distances), finding that various elastic parameters were affected by disorder but did not calculate microstructural parameters.  Zhang\cite{Zhang2018Thermal} calculated the effect of disorder on thermal transport in VT using two different models, finding opposite results with the classical theory predicting an increased in conductivity with disorder.  Even less is known about the microstructural factors of centroidal VT, which are VT where the mass centroid of each cell coincides with the generator point.  Wang calculated that CVT walls have thermal transport between the bounds, consistent with finite $\zeta$ but this parameter was not determined.  There appears to be a lack of high quality analysis needed for predicting the transport in these structures, which we can now address.

\section{\label{sec:Generation}Generation of CVT}
I now briefly outline how Centroidal Voronoi Tessellations may generated.  Several algorithms have been developed for construction\cite{Du1999Centroidal} of CVT, but we use Lloyd's method which simply consists of iteratively determining the VT of the centroids (but is slowly converging for large cell numbers).  In this article, initial generators were randomly distributed in a periodic triangular supercell (which seems more natural than the square supercell used in previous periodic CVT \cite{Zhang2012}).   Cell configurations can be characterized by the total inertia $J$ of the bounding polygons \cite{Du1999Centroidal} (compared the inertia $J_0$ of discs with the same area), which is a measure of compactness.  In general this correlates with irregularity, but even semi-regular lattices are spread out along this scale.  Calculations on semi-regular lattices show that this measure has the same ordering as the $\zeta$ parameter (seen below).

The number of generators in the unit cell has a strong effect on the possible random-seeded PCVT.  For example, exact hexagonal packing is only possible in triangular lattices for $m^2+mn+n^2$ (1,3,4,7,9,12,13,16,19,...) generators.  Low generators favour less optimal geometries: 5 gives 467 and 568 configurations and 6 admits irregular hexagons and 57 configurations.  From 8 onwards, irregular configurations are typically 567, sometimes occurring at generator numbers that also allow regular hexagons and occasionally irregular hexagons.  19 appears to be the last pure hexagonal configuration.  Due to the ubiquity of 567 configurations, this is a representative model for irregular CVT.  Regularity generally improves with increasing number of generators in the unit cell.

\section{\label{sec:Regular}Transport Parameters of Regular Lattices}
\begin{figure}[h]
\includegraphics[width=\columnwidth]{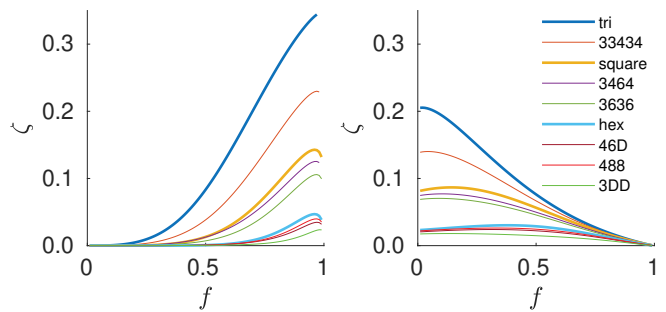}
\caption{\label{fig:RegularZeta} Summary of third order parameter for regular and selected semi-regular lattices as a function of fill factor, for (left) conformal and (right) polygonal inclusion shapes.  Additional microstructural parameters for these lattices are presented in the appendix.}
\end{figure}

To confirm the validity of this method, I calculated microstructural factors of conformal contours and polygonal prisms in regular and isotropic semi-regular arrays, which complements my previous survey of the fundamental resonance of these structures \cite{Arnold2017}.  The conformal contours are determined via a Schwartz-Christoffel transform \cite{Driscoll1996algorithm}.  A subset of these results concur with our previous work \cite{Gali2019}, which were in turn validated against other methods referenced therein. Note that the ordering of the lattice third-order parameter (3.12.12 $>$ 4.6.12 $\sim$ 4.8.8 $>$ 6 $>$ 3.6.3.6 $>$ 4 $\sim$ 3.4.6.4 $>$ 3.3.4.3.4 $>$ 3), as seen in Fig.~\ref{fig:RegularZeta}, is consistent with local symmetry and with the relative deviation in the fundamental resonance seen earlier \cite{Arnold2017}.  Further results up to 7\textsuperscript{th} order (Fig.~\ref{fig:HighOrder}), including circular inclusions in regular lattices (Fig.~\ref{fig:Discs}), are shown in the appendix.

Hyun and Torquato estimated the third order parameter (for sharp triangles and hexagons) \cite{Hyun2000}, but their results did not converge to the expected limit ($\zeta|_{f=1}=0.2043$) \cite{Hetherington1992} at high fill factor, possibly due to numerical problems.  They extracted the third order parameter by comparison of the effective modulus with the bounds in the limit of zero contrast \cite{Eischen1993}.  However, this approach can be problematic due to numerical instability at low contrast, which appears to be particularly acute in the case of triangles.  I performed a similar procedure using a commercial electrostatic FEM with much better sampling than the cited reference and carefully inspected the limiting behaviour, finding good agreement with our other results.

\section{\label{sec:VT}Transport Parameters of Voronoi Tesselations}

Finally, effective transport properties of these structures, with both conformal and polygonal inclusions, are surveyed.

\begin{figure}[h]
\includegraphics[width=\columnwidth]{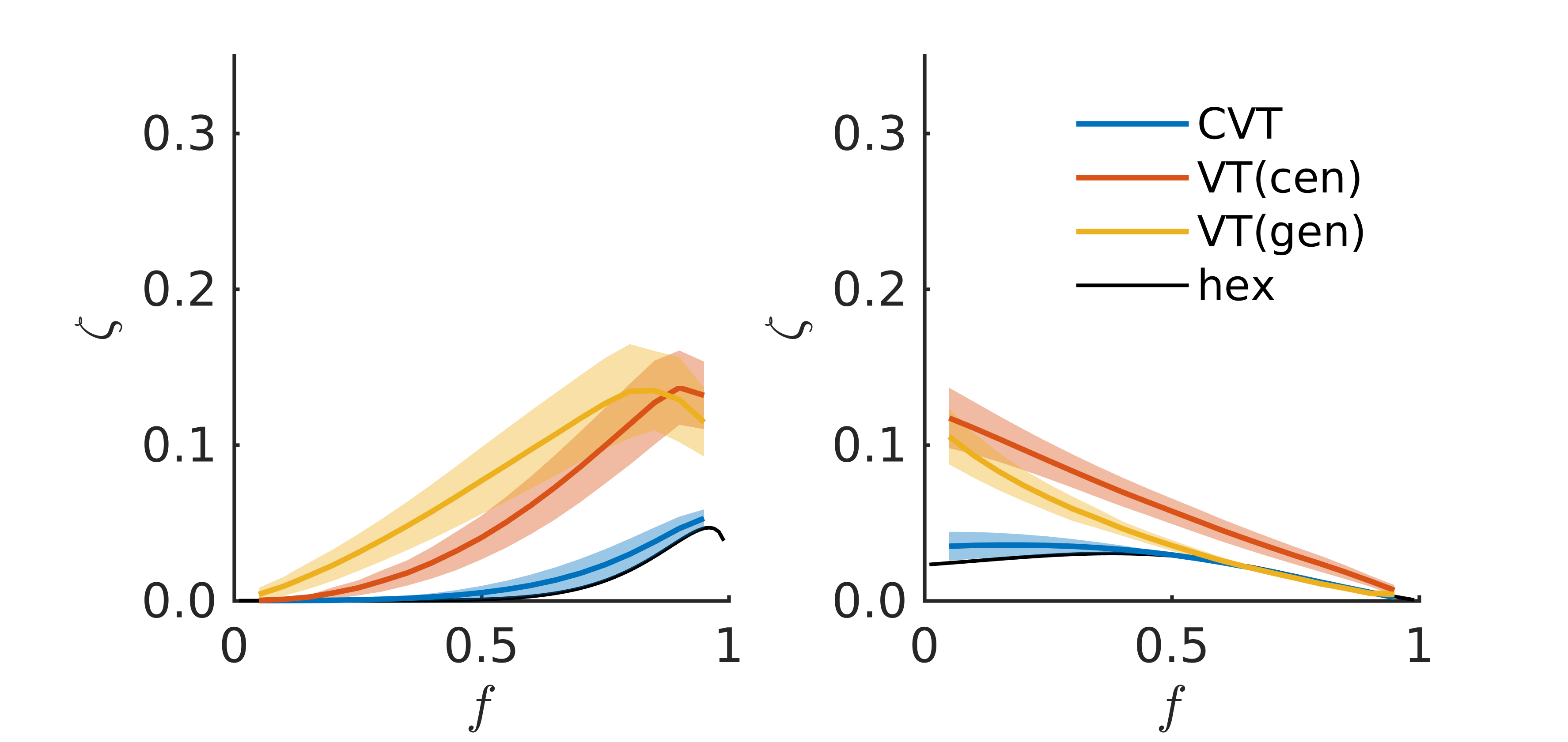}
\caption{\label{fig:VoronoiZeta} Effect of fill factor on the third order parameter for CVT and VT with inclusions (left:conformal, right:polygonal) evolving from centroids and generators.  The mean is shown as a solid thin line and the standard deviation is indicated by the transparent band.  Regular hexagonal lattices are shown for comparison.  Note that individual structures cover an arbitrary range of compactness.}
\end{figure}

Figure \ref{fig:VoronoiZeta} shows that $\zeta$ for CVT is not much higher than regular hexagonal lattices at large $\zeta$, which is perhaps not surprising given the significant proportion of hexagons ($\zeta|_{f=1}=0.023010$), together with the slightly stronger influence of pentagons ($\zeta|_{f=1}=0.040310$) compared to heptagons ($\zeta|_{f=1}=0.014371$).  Random VT (with inclusions evolving from either the cell centroids or the generators) have higher $\zeta$ than the more ordered CVT.  Analysis of compactness at a given fill-factor yields further insights below. 

\begin{figure}[h]
\includegraphics[width=\columnwidth]{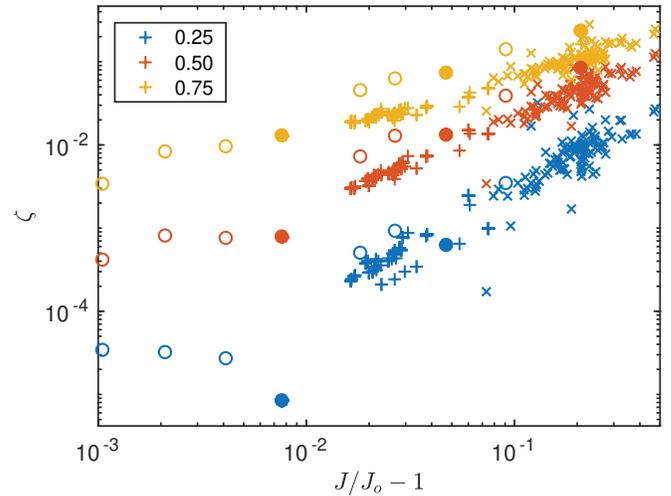}
\caption{\label{fig:ConformalSpread} Effect of compactness (cell inertia compared to a disc) on third order parameter for confocal inclusions.  Filled circles regular, open circles semi-regular, + crosses CVT, x crosses random VT (evolving from centroids).  Fill-factors 0.25, 0.50, 0.75.  The ordering of the (semi) regular lattices from left to right is 3.12.12, 4.8.8, 4.6.12, hexagons, $(3.6)^2$, 3.4.6.4, squares, $3^2.4.3$, triangles.}
\end{figure}

\begin{figure}[h]
\includegraphics[width=\columnwidth]{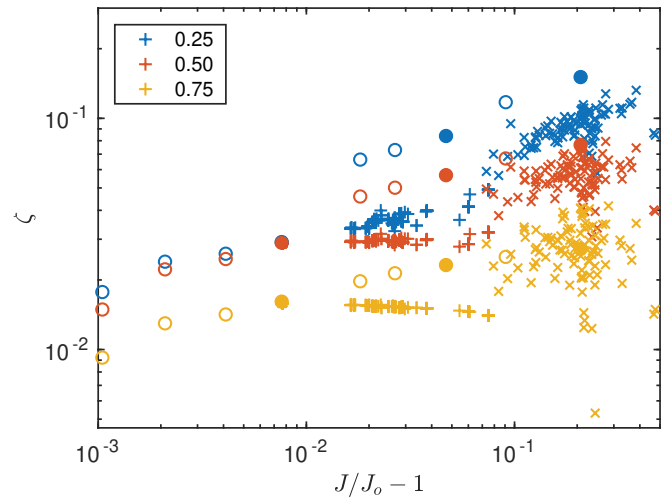}
\caption{\label{fig:PolygonalSpread} Effect of compactness on third order parameter for polygonal inclusions.  Filled circles regular, open circles semi-regular, + crosses CVT, x crosses random VT (evolving from centroids).  Fill-factors 0.25, 0.50, 0.75.}
\end{figure}

The main conclusion from Figs~\ref{fig:ConformalSpread} and \ref{fig:PolygonalSpread} is that CVT generally have lower $\zeta$ than comparable semi-regular results (3.6.3.6, 3.4.6.4) at a given compactness, especially for polygonal inclusions at high fill-factor.  Interestingly, the random VT in Fig.~\ref{fig:PolygonalSpread} are also closer to the bounds than comparable lattices (e.g. triangles).   There is a strong correlation between compactness and $\zeta$ for conformal inclusions (Fig.~\ref{fig:ConformalSpread}), but polygonal inclusions in CVT (Fig.~\ref{fig:PolygonalSpread}) have little correlation to compactness.

\section{\label{sec:Conclusion}Conclusion}
This article outlined a boundary element method for calculating the microstructural parameters to high accuracy, and calculated these values for regular and selected semi-regular lattices with both conformal and polygonal inclusions.  The results showed that previously reported third order values for triangular inclusions lattices are inaccurate.  Overall, centroidal VT have smaller third order parameter than less ordered VT.  CVT with conformal inclusions have third order values somewhat correlated to the relative inertial moment of the cells, and in many cases closer to the extremes than comparable semi-regular lattices.  This is probably due to the predominance of nearly hexagonal cells, which are more optimal than the triangles in nearby semi-regular lattices.  Overall, these findings helps to quantify a long held heuristic that low order bounds are an adequate description for random structures if they tend towards close-packed, but less so if disorder becomes large.

\begin{acknowledgments}
Helpful discussions with Marc Gali are acknowledged.
\end{acknowledgments}

\appendix*

\section{}
High order microstructural parameters are presented for polygonal, conformal and circular inclusions.

% \begin{figure*} % for wide
\begin{figure}[h]
%\begin{subfigure}
\includegraphics[width=\columnwidth]{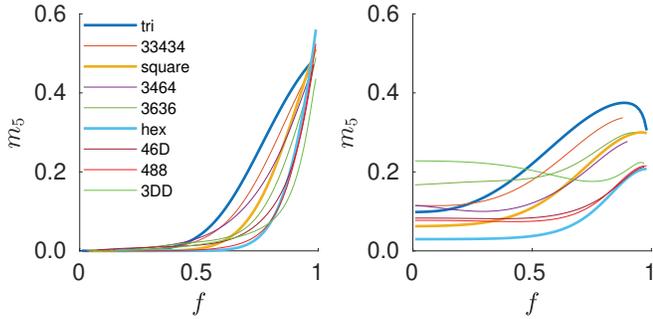}
\label{fig:Order5}
%\end{subfigure}
\caption{\label{fig:HighOrder} Fifth order microstructural parameters of (left) conformal and (right) polygonal prisms in regular and isotropic semi-regular lattices.  These results are converged to better than 0.01, calculated using $\sim 10^4$ boundary points.}
\end{figure}
\begin{figure}[h]
%\begin{subfigure}
\includegraphics[width=\columnwidth]{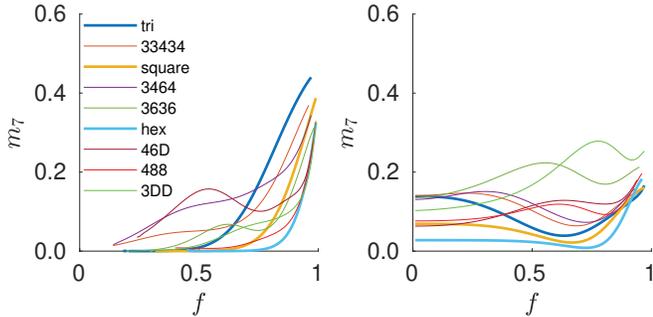}
\label{fig:Order7}
%\end{subfigure}
\caption{\label{fig:HighOrder} Seventh order microstructural parameters of (left) conformal and (right) polygonal prisms in regular and isotropic semi-regular lattices.}
\end{figure}

\begin{figure}[h]
\centering

\hfill\break
\includegraphics[width=\columnwidth]{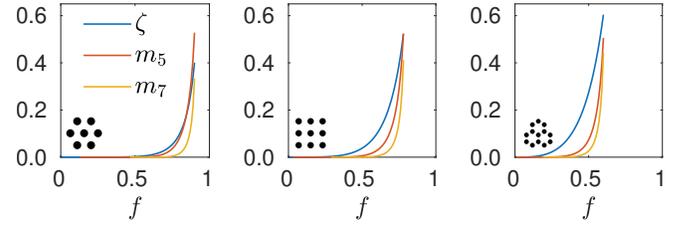}
\caption{\label{fig:Discs} Microstructural parameters of circular inclusions in (left) hexagonal, (middle) square and (right) triangular lattices.  Only fill factors below the percolation threshold are shown.}
\end{figure}

%If there is only one appendix, then the letter ``A'' should not
%appear. This is suppressed by using the star version of the appendix
%command (\verb+\appendix*+ in the place of \verb+\appendix+).

%\nocite{*}

\bibliography{BEMVoro}% Produces the bibliography via BibTeX.

\providecommand{\noopsort}[1]{}\providecommand{\singleletter}[1]{#1}%
\begin{thebibliography}{35}%
\makeatletter
\providecommand \@ifxundefined [1]{%
 \@ifx{#1\undefined}
}%
\providecommand \@ifnum [1]{%
 \ifnum #1\expandafter \@firstoftwo
 \else \expandafter \@secondoftwo
 \fi
}%
\providecommand \@ifx [1]{%
 \ifx #1\expandafter \@firstoftwo
 \else \expandafter \@secondoftwo
 \fi
}%
\providecommand \natexlab [1]{#1}%
\providecommand \enquote  [1]{``#1''}%
\providecommand \bibnamefont  [1]{#1}%
\providecommand \bibfnamefont [1]{#1}%
\providecommand \citenamefont [1]{#1}%
\providecommand \href@noop [0]{\@secondoftwo}%
\providecommand \href [0]{\begingroup \@sanitize@url \@href}%
\providecommand \@href[1]{\@@startlink{#1}\@@href}%
\providecommand \@@href[1]{\endgroup#1\@@endlink}%
\providecommand \@sanitize@url [0]{\catcode `\\12\catcode `\$12\catcode
  `\&12\catcode `\#12\catcode `\^12\catcode `\_12\catcode `\%12\relax}%
\providecommand \@@startlink[1]{}%
\providecommand \@@endlink[0]{}%
\providecommand \url  [0]{\begingroup\@sanitize@url \@url }%
\providecommand \@url [1]{\endgroup\@href {#1}{\urlprefix }}%
\providecommand \urlprefix  [0]{URL }%
\providecommand \Eprint [0]{\href }%
\providecommand \doibase [0]{http://dx.doi.org/}%
\providecommand \selectlanguage [0]{\@gobble}%
\providecommand \bibinfo  [0]{\@secondoftwo}%
\providecommand \bibfield  [0]{\@secondoftwo}%
\providecommand \translation [1]{[#1]}%
\providecommand \BibitemOpen [0]{}%
\providecommand \bibitemStop [0]{}%
\providecommand \bibitemNoStop [0]{.\EOS\space}%
\providecommand \EOS [0]{\spacefactor3000\relax}%
\providecommand \BibitemShut  [1]{\csname bibitem#1\endcsname}%
\let\auto@bib@innerbib\@empty
%</preamble>
\bibitem [{\citenamefont {Torquato}(2002)}]{Torquato2002}%
  \BibitemOpen
  \bibfield  {author} {\bibinfo {author} {\bibfnamefont {S.}~\bibnamefont
  {Torquato}},\ }\href@noop {} {\emph {\bibinfo {title} {{Random Heterogeneous
  Materials: Microstructure and Macroscopic Properties}}}}\ (\bibinfo
  {publisher} {Springer-Verlag},\ \bibinfo {address} {New York},\ \bibinfo
  {year} {2002})\BibitemShut {NoStop}%
\bibitem [{\citenamefont {Milton}(2002)}]{Milton2002}%
  \BibitemOpen
  \bibfield  {author} {\bibinfo {author} {\bibfnamefont {G.~W.}\ \bibnamefont
  {Milton}},\ }\href@noop {} {\emph {\bibinfo {title} {{The Theory of
  Composites}}}}\ (\bibinfo  {publisher} {Cambridge University Press},\
  \bibinfo {year} {2002})\BibitemShut {NoStop}%
\bibitem [{\citenamefont {Roberts}\ and\ \citenamefont
  {Teubner}(1995)}]{Roberts1995}%
  \BibitemOpen
  \bibfield  {author} {\bibinfo {author} {\bibfnamefont {A.~P.}\ \bibnamefont
  {Roberts}}\ and\ \bibinfo {author} {\bibfnamefont {M.}~\bibnamefont
  {Teubner}},\ }\bibfield  {title} {\enquote {\bibinfo {title} {{Transport
  properties of heterogeneous materials derived from Gaussian random fields:
  Bonds and simulation}},}\ }\href@noop {} {\bibfield  {journal} {\bibinfo
  {journal} {Physical Review E}\ }\textbf {\bibinfo {volume} {51}},\ \bibinfo
  {pages} {4141--4154} (\bibinfo {year} {1995})}\BibitemShut {NoStop}%
\bibitem [{\citenamefont {Winterfeld}, \citenamefont {Scriven},\ and\
  \citenamefont {Davis}(1981)}]{Winterfeld1981percolation}%
  \BibitemOpen
  \bibfield  {author} {\bibinfo {author} {\bibfnamefont {P.}~\bibnamefont
  {Winterfeld}}, \bibinfo {author} {\bibfnamefont {L.}~\bibnamefont {Scriven}},
  \ and\ \bibinfo {author} {\bibfnamefont {H.}~\bibnamefont {Davis}},\
  }\bibfield  {title} {\enquote {\bibinfo {title} {Percolation and conductivity
  of random two-dimensional composites},}\ }\href@noop {} {\bibfield  {journal}
  {\bibinfo  {journal} {Journal of Physics C: Solid State Physics}\ }\textbf
  {\bibinfo {volume} {14}},\ \bibinfo {pages} {2361} (\bibinfo {year}
  {1981})}\BibitemShut {NoStop}%
\bibitem [{\citenamefont {Jerauld}\ \emph {et~al.}(1984)\citenamefont
  {Jerauld}, \citenamefont {Hatfield}, \citenamefont {Scriven},\ and\
  \citenamefont {Davis}}]{Jerauld1984percolation}%
  \BibitemOpen
  \bibfield  {author} {\bibinfo {author} {\bibfnamefont {G.}~\bibnamefont
  {Jerauld}}, \bibinfo {author} {\bibfnamefont {J.}~\bibnamefont {Hatfield}},
  \bibinfo {author} {\bibfnamefont {L.}~\bibnamefont {Scriven}}, \ and\
  \bibinfo {author} {\bibfnamefont {H.}~\bibnamefont {Davis}},\ }\bibfield
  {title} {\enquote {\bibinfo {title} {Percolation and conduction on voronoi
  and triangular networks: a case study in topological disorder},}\ }\href@noop
  {} {\bibfield  {journal} {\bibinfo  {journal} {Journal of Physics C: Solid
  State Physics}\ }\textbf {\bibinfo {volume} {17}},\ \bibinfo {pages} {1519}
  (\bibinfo {year} {1984})}\BibitemShut {NoStop}%
\bibitem [{\citenamefont {Torquato}\ and\ \citenamefont
  {Stillinger}(2003)}]{TorquatoStillinger2003}%
  \BibitemOpen
  \bibfield  {author} {\bibinfo {author} {\bibfnamefont {S.}~\bibnamefont
  {Torquato}}\ and\ \bibinfo {author} {\bibfnamefont {F.~H.}\ \bibnamefont
  {Stillinger}},\ }\bibfield  {title} {\enquote {\bibinfo {title} {Local
  density fluctuations, hyperuniformity, and order metrics},}\ }\href@noop {}
  {\bibfield  {journal} {\bibinfo  {journal} {Physical Review E}\ }\textbf
  {\bibinfo {volume} {68}},\ \bibinfo {pages} {041113} (\bibinfo {year}
  {2003})}\BibitemShut {NoStop}%
\bibitem [{\citenamefont {Man}\ \emph {et~al.}(2013)\citenamefont {Man},
  \citenamefont {Florescu}, \citenamefont {Williamson}, \citenamefont {He},
  \citenamefont {Hashemizad}, \citenamefont {Leung}, \citenamefont {Liner},
  \citenamefont {Torquato}, \citenamefont {Chaikin},\ and\ \citenamefont
  {Steinhardt}}]{Man2013isotropic}%
  \BibitemOpen
  \bibfield  {author} {\bibinfo {author} {\bibfnamefont {W.}~\bibnamefont
  {Man}}, \bibinfo {author} {\bibfnamefont {M.}~\bibnamefont {Florescu}},
  \bibinfo {author} {\bibfnamefont {E.~P.}\ \bibnamefont {Williamson}},
  \bibinfo {author} {\bibfnamefont {Y.}~\bibnamefont {He}}, \bibinfo {author}
  {\bibfnamefont {S.~R.}\ \bibnamefont {Hashemizad}}, \bibinfo {author}
  {\bibfnamefont {B.~Y.}\ \bibnamefont {Leung}}, \bibinfo {author}
  {\bibfnamefont {D.~R.}\ \bibnamefont {Liner}}, \bibinfo {author}
  {\bibfnamefont {S.}~\bibnamefont {Torquato}}, \bibinfo {author}
  {\bibfnamefont {P.~M.}\ \bibnamefont {Chaikin}}, \ and\ \bibinfo {author}
  {\bibfnamefont {P.~J.}\ \bibnamefont {Steinhardt}},\ }\bibfield  {title}
  {\enquote {\bibinfo {title} {Isotropic band gaps and freeform waveguides
  observed in hyperuniform disordered photonic solids},}\ }\href@noop {}
  {\bibfield  {journal} {\bibinfo  {journal} {Proceedings of the National
  Academy of Sciences}\ }\textbf {\bibinfo {volume} {110}},\ \bibinfo {pages}
  {15886--15891} (\bibinfo {year} {2013})}\BibitemShut {NoStop}%
\bibitem [{\citenamefont {Liu}\ \emph {et~al.}(2018)\citenamefont {Liu},
  \citenamefont {Zhao}, \citenamefont {Wang},\ and\ \citenamefont
  {Fang}}]{Liu2018absorption}%
  \BibitemOpen
  \bibfield  {author} {\bibinfo {author} {\bibfnamefont {M.}~\bibnamefont
  {Liu}}, \bibinfo {author} {\bibfnamefont {C.}~\bibnamefont {Zhao}}, \bibinfo
  {author} {\bibfnamefont {B.}~\bibnamefont {Wang}}, \ and\ \bibinfo {author}
  {\bibfnamefont {X.}~\bibnamefont {Fang}},\ }\bibfield  {title} {\enquote
  {\bibinfo {title} {Role of short-range order in manipulating light absorption
  in disordered media},}\ }\href@noop {} {\bibfield  {journal} {\bibinfo
  {journal} {JOSA B}\ }\textbf {\bibinfo {volume} {35}},\ \bibinfo {pages}
  {504--513} (\bibinfo {year} {2018})}\BibitemShut {NoStop}%
\bibitem [{\citenamefont {Florescu}, \citenamefont {Torquato},\ and\
  \citenamefont {Steinhardt}(2009)}]{Florescu2009designer}%
  \BibitemOpen
  \bibfield  {author} {\bibinfo {author} {\bibfnamefont {M.}~\bibnamefont
  {Florescu}}, \bibinfo {author} {\bibfnamefont {S.}~\bibnamefont {Torquato}},
  \ and\ \bibinfo {author} {\bibfnamefont {P.~J.}\ \bibnamefont {Steinhardt}},\
  }\bibfield  {title} {\enquote {\bibinfo {title} {Designer disordered
  materials with large, complete photonic band gaps},}\ }\href@noop {}
  {\bibfield  {journal} {\bibinfo  {journal} {Proceedings of the National
  Academy of Sciences}\ }\textbf {\bibinfo {volume} {106}},\ \bibinfo {pages}
  {20658--20663} (\bibinfo {year} {2009})}\BibitemShut {NoStop}%
\bibitem [{\citenamefont {Du}, \citenamefont {Faber},\ and\ \citenamefont
  {Gunzburger}(1999)}]{Du1999Centroidal}%
  \BibitemOpen
  \bibfield  {author} {\bibinfo {author} {\bibfnamefont {Q.}~\bibnamefont
  {Du}}, \bibinfo {author} {\bibfnamefont {V.}~\bibnamefont {Faber}}, \ and\
  \bibinfo {author} {\bibfnamefont {M.}~\bibnamefont {Gunzburger}},\ }\bibfield
   {title} {\enquote {\bibinfo {title} {Centroidal voronoi tessellations:
  Applications and algorithms},}\ }\href@noop {} {\bibfield  {journal}
  {\bibinfo  {journal} {SIAM review}\ }\textbf {\bibinfo {volume} {41}},\
  \bibinfo {pages} {637--676} (\bibinfo {year} {1999})}\BibitemShut {NoStop}%
\bibitem [{\citenamefont {Klatt}\ \emph {et~al.}(2019)\citenamefont {Klatt},
  \citenamefont {Lovri{\'c}}, \citenamefont {Chen}, \citenamefont {Kapfer},
  \citenamefont {Schaller}, \citenamefont {Sch{\"o}nh{\"o}fer}, \citenamefont
  {Gardiner}, \citenamefont {Smith}, \citenamefont {Schr{\"o}der-Turk},\ and\
  \citenamefont {Torquato}}]{Klatt2019}%
  \BibitemOpen
  \bibfield  {author} {\bibinfo {author} {\bibfnamefont {M.~A.}\ \bibnamefont
  {Klatt}}, \bibinfo {author} {\bibfnamefont {J.}~\bibnamefont {Lovri{\'c}}},
  \bibinfo {author} {\bibfnamefont {D.}~\bibnamefont {Chen}}, \bibinfo {author}
  {\bibfnamefont {S.~C.}\ \bibnamefont {Kapfer}}, \bibinfo {author}
  {\bibfnamefont {F.~M.}\ \bibnamefont {Schaller}}, \bibinfo {author}
  {\bibfnamefont {P.~W.}\ \bibnamefont {Sch{\"o}nh{\"o}fer}}, \bibinfo {author}
  {\bibfnamefont {B.~S.}\ \bibnamefont {Gardiner}}, \bibinfo {author}
  {\bibfnamefont {A.-S.}\ \bibnamefont {Smith}}, \bibinfo {author}
  {\bibfnamefont {G.~E.}\ \bibnamefont {Schr{\"o}der-Turk}}, \ and\ \bibinfo
  {author} {\bibfnamefont {S.}~\bibnamefont {Torquato}},\ }\bibfield  {title}
  {\enquote {\bibinfo {title} {Universal hidden order in amorphous cellular
  geometries},}\ }\href@noop {} {\bibfield  {journal} {\bibinfo  {journal}
  {Nature communications}\ }\textbf {\bibinfo {volume} {10}},\ \bibinfo {pages}
  {811} (\bibinfo {year} {2019})}\BibitemShut {NoStop}%
\bibitem [{\citenamefont {Previdi}\ \emph {et~al.}(2019)\citenamefont
  {Previdi}, \citenamefont {Levchenko}, \citenamefont {Arnold}, \citenamefont
  {Gali}, \citenamefont {Bazaka}, \citenamefont {Xu}, \citenamefont {Ostrikov},
  \citenamefont {Bray}, \citenamefont {Jin},\ and\ \citenamefont
  {Fang}}]{Previdi2019}%
  \BibitemOpen
  \bibfield  {author} {\bibinfo {author} {\bibfnamefont {R.}~\bibnamefont
  {Previdi}}, \bibinfo {author} {\bibfnamefont {I.}~\bibnamefont {Levchenko}},
  \bibinfo {author} {\bibfnamefont {M.}~\bibnamefont {Arnold}}, \bibinfo
  {author} {\bibfnamefont {M.}~\bibnamefont {Gali}}, \bibinfo {author}
  {\bibfnamefont {K.}~\bibnamefont {Bazaka}}, \bibinfo {author} {\bibfnamefont
  {S.}~\bibnamefont {Xu}}, \bibinfo {author} {\bibfnamefont {K.~K.}\
  \bibnamefont {Ostrikov}}, \bibinfo {author} {\bibfnamefont {K.}~\bibnamefont
  {Bray}}, \bibinfo {author} {\bibfnamefont {D.}~\bibnamefont {Jin}}, \ and\
  \bibinfo {author} {\bibfnamefont {J.}~\bibnamefont {Fang}},\ }\bibfield
  {title} {\enquote {\bibinfo {title} {Plasmonic platform based on nanoporous
  alumina membranes: order control via self-assembly},}\ }\href@noop {}
  {\bibfield  {journal} {\bibinfo  {journal} {Journal of Materials Chemistry
  A}\ }\textbf {\bibinfo {volume} {7}},\ \bibinfo {pages} {9565--9577}
  (\bibinfo {year} {2019})}\BibitemShut {NoStop}%
\bibitem [{\citenamefont {Torquato}\ \emph {et~al.}(1998)\citenamefont
  {Torquato}, \citenamefont {Gibiansky}, \citenamefont {Silva},\ and\
  \citenamefont {Gibson}}]{TorquatoGibiansky1998}%
  \BibitemOpen
  \bibfield  {author} {\bibinfo {author} {\bibfnamefont {S.}~\bibnamefont
  {Torquato}}, \bibinfo {author} {\bibfnamefont {L.}~\bibnamefont {Gibiansky}},
  \bibinfo {author} {\bibfnamefont {M.}~\bibnamefont {Silva}}, \ and\ \bibinfo
  {author} {\bibfnamefont {L.}~\bibnamefont {Gibson}},\ }\bibfield  {title}
  {\enquote {\bibinfo {title} {Effective mechanical and transport properties of
  cellular solids},}\ }\href@noop {} {\bibfield  {journal} {\bibinfo  {journal}
  {International Journal of Mechanical Sciences}\ }\textbf {\bibinfo {volume}
  {40}},\ \bibinfo {pages} {71--82} (\bibinfo {year} {1998})}\BibitemShut
  {NoStop}%
\bibitem [{\citenamefont {Vrettos}, \citenamefont {Imakoma},\ and\
  \citenamefont {Okazaki}(1989)}]{Vrettos1989effective}%
  \BibitemOpen
  \bibfield  {author} {\bibinfo {author} {\bibfnamefont {N.~A.}\ \bibnamefont
  {Vrettos}}, \bibinfo {author} {\bibfnamefont {H.}~\bibnamefont {Imakoma}}, \
  and\ \bibinfo {author} {\bibfnamefont {M.}~\bibnamefont {Okazaki}},\
  }\bibfield  {title} {\enquote {\bibinfo {title} {An effective medium
  treatment of the transport properties of a voronoi tesselated network},}\
  }\href@noop {} {\bibfield  {journal} {\bibinfo  {journal} {Journal of Applied
  Physics}\ }\textbf {\bibinfo {volume} {66}},\ \bibinfo {pages} {2873--2878}
  (\bibinfo {year} {1989})}\BibitemShut {NoStop}%
\bibitem [{\citenamefont {Torquato}\ and\ \citenamefont
  {Lado}(1992)}]{TorquatoLado1992}%
  \BibitemOpen
  \bibfield  {author} {\bibinfo {author} {\bibfnamefont {S.}~\bibnamefont
  {Torquato}}\ and\ \bibinfo {author} {\bibfnamefont {F.}~\bibnamefont
  {Lado}},\ }\bibfield  {title} {\enquote {\bibinfo {title} {Improved bounds on
  the effective elastic moduli of random arrays of cylinders},}\ }\href@noop {}
  {\bibfield  {journal} {\bibinfo  {journal} {Journal of Applied Mechanics}\
  }\textbf {\bibinfo {volume} {59}},\ \bibinfo {pages} {1--6} (\bibinfo {year}
  {1992})}\BibitemShut {NoStop}%
\bibitem [{\citenamefont {Arnold}(2017)}]{Arnold2017}%
  \BibitemOpen
  \bibfield  {author} {\bibinfo {author} {\bibfnamefont {M.~D.}\ \bibnamefont
  {Arnold}},\ }\bibfield  {title} {\enquote {\bibinfo {title} {{Single-mode
  tuning of the plasmon resonance in high-density pillar arrays}},}\ }\href
  {\doibase 10.1088/1361-648X/aa57c8} {\bibfield  {journal} {\bibinfo
  {journal} {Journal of Physics Condensed Matter}\ }\textbf {\bibinfo {volume}
  {29}} (\bibinfo {year} {2017}),\ 10.1088/1361-648X/aa57c8}\BibitemShut
  {NoStop}%
\bibitem [{\citenamefont {Vigdergauz}(1994)}]{Vigdergauz1994two}%
  \BibitemOpen
  \bibfield  {author} {\bibinfo {author} {\bibfnamefont {S.}~\bibnamefont
  {Vigdergauz}},\ }\bibfield  {title} {\enquote {\bibinfo {title}
  {Two-dimensional grained composites of extreme rigidity},}\ }\href@noop {}
  {\bibfield  {journal} {\bibinfo  {journal} {Journal of applied mechanics}\
  }\textbf {\bibinfo {volume} {61}},\ \bibinfo {pages} {390--394} (\bibinfo
  {year} {1994})}\BibitemShut {NoStop}%
\bibitem [{\citenamefont {Vigdergauz}(1996)}]{Vigdergauz1996rhombic}%
  \BibitemOpen
  \bibfield  {author} {\bibinfo {author} {\bibfnamefont {S.}~\bibnamefont
  {Vigdergauz}},\ }\bibfield  {title} {\enquote {\bibinfo {title} {Rhombic
  lattice of equi-stress inclusions in an elastic plate},}\ }\href@noop {}
  {\bibfield  {journal} {\bibinfo  {journal} {Quarterly journal of mechanics
  and applied mathematics}\ }\textbf {\bibinfo {volume} {49}},\ \bibinfo
  {pages} {565--580} (\bibinfo {year} {1996})}\BibitemShut {NoStop}%
\bibitem [{\citenamefont {Vigdergauz}(1999)}]{Vigdergauz1999energy}%
  \BibitemOpen
  \bibfield  {author} {\bibinfo {author} {\bibfnamefont {S.}~\bibnamefont
  {Vigdergauz}},\ }\bibfield  {title} {\enquote {\bibinfo {title}
  {Energy-minimizing inclusions in a planar elastic structure with
  macroisotropy},}\ }\href@noop {} {\bibfield  {journal} {\bibinfo  {journal}
  {Structural optimization}\ }\textbf {\bibinfo {volume} {17}},\ \bibinfo
  {pages} {104--112} (\bibinfo {year} {1999})}\BibitemShut {NoStop}%
\bibitem [{\citenamefont {Hyun}\ and\ \citenamefont
  {Torquato}(2002)}]{hyun2002optimal}%
  \BibitemOpen
  \bibfield  {author} {\bibinfo {author} {\bibfnamefont {S.}~\bibnamefont
  {Hyun}}\ and\ \bibinfo {author} {\bibfnamefont {S.}~\bibnamefont
  {Torquato}},\ }\bibfield  {title} {\enquote {\bibinfo {title} {Optimal and
  manufacturable two-dimensional, kagome-like cellular solids},}\ }\href@noop
  {} {\bibfield  {journal} {\bibinfo  {journal} {Journal of Materials
  Research}\ }\textbf {\bibinfo {volume} {17}},\ \bibinfo {pages} {137--144}
  (\bibinfo {year} {2002})}\BibitemShut {NoStop}%
\bibitem [{\citenamefont {Driscoll}(1996)}]{Driscoll1996algorithm}%
  \BibitemOpen
  \bibfield  {author} {\bibinfo {author} {\bibfnamefont {T.~A.}\ \bibnamefont
  {Driscoll}},\ }\bibfield  {title} {\enquote {\bibinfo {title} {Algorithm 756:
  A matlab toolbox for schwarz-christoffel mapping},}\ }\href@noop {}
  {\bibfield  {journal} {\bibinfo  {journal} {ACM Transactions on Mathematical
  Software (TOMS)}\ }\textbf {\bibinfo {volume} {22}},\ \bibinfo {pages}
  {168--186} (\bibinfo {year} {1996})}\BibitemShut {NoStop}%
\bibitem [{\citenamefont {Bergman}(1979)}]{Bergman1979}%
  \BibitemOpen
  \bibfield  {author} {\bibinfo {author} {\bibfnamefont {D.~J.}\ \bibnamefont
  {Bergman}},\ }\bibfield  {title} {\enquote {\bibinfo {title} {{Dielectric
  constant of a two-component granular composite: A practical scheme for
  calculating the pole spectrum}},}\ }\href {\doibase 10.1103/PhysRevB.19.2359}
  {\bibfield  {journal} {\bibinfo  {journal} {Physical Review B}\ }\textbf
  {\bibinfo {volume} {19}},\ \bibinfo {pages} {2359--2368} (\bibinfo {year}
  {1979})}\BibitemShut {NoStop}%
\bibitem [{\citenamefont {Djordjevi{\'c}}, \citenamefont {Hetherington},\ and\
  \citenamefont {Thorpe}(1996)}]{Djordjevic1996spectral}%
  \BibitemOpen
  \bibfield  {author} {\bibinfo {author} {\bibfnamefont {B.}~\bibnamefont
  {Djordjevi{\'c}}}, \bibinfo {author} {\bibfnamefont {J.}~\bibnamefont
  {Hetherington}}, \ and\ \bibinfo {author} {\bibfnamefont {M.}~\bibnamefont
  {Thorpe}},\ }\bibfield  {title} {\enquote {\bibinfo {title} {Spectral
  function for a conducting sheet containing circular inclusions},}\
  }\href@noop {} {\bibfield  {journal} {\bibinfo  {journal} {Physical Review
  B}\ }\textbf {\bibinfo {volume} {53}},\ \bibinfo {pages} {14862} (\bibinfo
  {year} {1996})}\BibitemShut {NoStop}%
\bibitem [{\citenamefont {Brown}(1955)}]{Brown1955}%
  \BibitemOpen
  \bibfield  {author} {\bibinfo {author} {\bibfnamefont {W.~F.}\ \bibnamefont
  {Brown}},\ }\bibfield  {title} {\enquote {\bibinfo {title} {{Solid Mixture
  Permittivities}},}\ }\href {\doibase 10.1063/1.1742339} {\bibfield  {journal}
  {\bibinfo  {journal} {The Journal of Chemical Physics}\ }\textbf {\bibinfo
  {volume} {23}},\ \bibinfo {pages} {1514--1517} (\bibinfo {year}
  {1955})}\BibitemShut {NoStop}%
\bibitem [{\citenamefont {Sen}\ and\ \citenamefont {Torquato}(1989)}]{Sen1989}%
  \BibitemOpen
  \bibfield  {author} {\bibinfo {author} {\bibfnamefont {A.~K.}\ \bibnamefont
  {Sen}}\ and\ \bibinfo {author} {\bibfnamefont {S.}~\bibnamefont {Torquato}},\
  }\bibfield  {title} {\enquote {\bibinfo {title} {{Effective Conductivity of
  Anisotropic 2-Phase Composite Media}},}\ }\href@noop {} {\bibfield  {journal}
  {\bibinfo  {journal} {Physical Review B}\ }\textbf {\bibinfo {volume} {39}},\
  \bibinfo {pages} {4504--4515} (\bibinfo {year} {1989})}\BibitemShut {NoStop}%
\bibitem [{\citenamefont {Engstr{\"{o}}m}(2005)}]{Engstrom2005}%
  \BibitemOpen
  \bibfield  {author} {\bibinfo {author} {\bibfnamefont {C.}~\bibnamefont
  {Engstr{\"{o}}m}},\ }\bibfield  {title} {\enquote {\bibinfo {title} {{Bounds
  on the effective tensor and the structural parameters for anisotropic
  two-phase composite material}},}\ }\href {\doibase
  10.1088/0022-3727/38/19/019} {\bibfield  {journal} {\bibinfo  {journal}
  {Journal of Physics D: Applied Physics}\ }\textbf {\bibinfo {volume} {38}},\
  \bibinfo {pages} {3695--3702} (\bibinfo {year} {2005})}\BibitemShut {NoStop}%
\bibitem [{\citenamefont {Gal\'{\i}}\ and\ \citenamefont
  {Arnold}(2019)}]{Gali2019}%
  \BibitemOpen
  \bibfield  {author} {\bibinfo {author} {\bibfnamefont {M.~A.}\ \bibnamefont
  {Gal\'{\i}}}\ and\ \bibinfo {author} {\bibfnamefont {M.~D.}\ \bibnamefont
  {Arnold}},\ }\bibfield  {title} {\enquote {\bibinfo {title} {Recurrent
  approach to effective material properties with application to anisotropic
  binarized random fields},}\ }\href {\doibase 10.1103/PhysRevB.99.054210}
  {\bibfield  {journal} {\bibinfo  {journal} {Phys. Rev. B}\ }\textbf {\bibinfo
  {volume} {99}},\ \bibinfo {pages} {054210} (\bibinfo {year}
  {2019})}\BibitemShut {NoStop}%
\bibitem [{\citenamefont {Mayergoyz}, \citenamefont {Fredkin},\ and\
  \citenamefont {Zhang}(2005)}]{Mayergoyz2005}%
  \BibitemOpen
  \bibfield  {author} {\bibinfo {author} {\bibfnamefont {I.~D.}\ \bibnamefont
  {Mayergoyz}}, \bibinfo {author} {\bibfnamefont {D.~R.}\ \bibnamefont
  {Fredkin}}, \ and\ \bibinfo {author} {\bibfnamefont {Z.}~\bibnamefont
  {Zhang}},\ }\bibfield  {title} {\enquote {\bibinfo {title} {Electrostatic
  (plasmon) resonances in nanoparticles},}\ }\href@noop {} {\bibfield
  {journal} {\bibinfo  {journal} {Physical Review B}\ }\textbf {\bibinfo
  {volume} {72}},\ \bibinfo {pages} {155412} (\bibinfo {year}
  {2005})}\BibitemShut {NoStop}%
\bibitem [{\citenamefont {Jones}(1945)}]{Jones1945Ellipsoid}%
  \BibitemOpen
  \bibfield  {author} {\bibinfo {author} {\bibfnamefont {R.~C.}\ \bibnamefont
  {Jones}},\ }\bibfield  {title} {\enquote {\bibinfo {title} {A generalization
  of the dielectric ellipsoid problem},}\ }\href@noop {} {\bibfield  {journal}
  {\bibinfo  {journal} {Physical Review}\ }\textbf {\bibinfo {volume} {68}},\
  \bibinfo {pages} {93} (\bibinfo {year} {1945})}\BibitemShut {NoStop}%
\bibitem [{\citenamefont {Zhu}, \citenamefont {Hobdell},\ and\ \citenamefont
  {Windle}(2001)}]{Zhu2001Disorder}%
  \BibitemOpen
  \bibfield  {author} {\bibinfo {author} {\bibfnamefont {H.}~\bibnamefont
  {Zhu}}, \bibinfo {author} {\bibfnamefont {J.}~\bibnamefont {Hobdell}}, \ and\
  \bibinfo {author} {\bibfnamefont {A.}~\bibnamefont {Windle}},\ }\bibfield
  {title} {\enquote {\bibinfo {title} {Effects of cell irregularity on the
  elastic properties of 2d voronoi honeycombs},}\ }\href@noop {} {\bibfield
  {journal} {\bibinfo  {journal} {Journal of the Mechanics and Physics of
  Solids}\ }\textbf {\bibinfo {volume} {49}},\ \bibinfo {pages} {857--870}
  (\bibinfo {year} {2001})}\BibitemShut {NoStop}%
\bibitem [{\citenamefont {Zhang}(2018)}]{Zhang2018Thermal}%
  \BibitemOpen
  \bibfield  {author} {\bibinfo {author} {\bibfnamefont {J.}~\bibnamefont
  {Zhang}},\ }\bibfield  {title} {\enquote {\bibinfo {title} {Effects of cell
  irregularity on the thermal conductivity of carbon honeycombs},}\ }\href@noop
  {} {\bibfield  {journal} {\bibinfo  {journal} {Carbon}\ }\textbf {\bibinfo
  {volume} {131}},\ \bibinfo {pages} {127--136} (\bibinfo {year}
  {2018})}\BibitemShut {NoStop}%
\bibitem [{\citenamefont {Zhang}, \citenamefont {Emelianenko},\ and\
  \citenamefont {Du}(2012)}]{Zhang2012}%
  \BibitemOpen
  \bibfield  {author} {\bibinfo {author} {\bibfnamefont {J.}~\bibnamefont
  {Zhang}}, \bibinfo {author} {\bibfnamefont {M.}~\bibnamefont {Emelianenko}},
  \ and\ \bibinfo {author} {\bibfnamefont {Q.}~\bibnamefont {Du}},\ }\bibfield
  {title} {\enquote {\bibinfo {title} {Periodic centroidal voronoi
  tessellations.}}\ }\href@noop {} {\bibfield  {journal} {\bibinfo  {journal}
  {International Journal of Numerical Analysis \& Modeling}\ }\textbf {\bibinfo
  {volume} {9}} (\bibinfo {year} {2012})}\BibitemShut {NoStop}%
\bibitem [{\citenamefont {Hyun}\ and\ \citenamefont
  {Torquato}(2000)}]{Hyun2000}%
  \BibitemOpen
  \bibfield  {author} {\bibinfo {author} {\bibfnamefont {S.}~\bibnamefont
  {Hyun}}\ and\ \bibinfo {author} {\bibfnamefont {S.}~\bibnamefont
  {Torquato}},\ }\bibfield  {title} {\enquote {\bibinfo {title} {{Effective
  elastic and transport properties of regular honeycombs for all densities}},}\
  }\href {\doibase 10.1557/JMR.2000.0285} {\bibfield  {journal} {\bibinfo
  {journal} {Journal of Materials Research}\ }\textbf {\bibinfo {volume}
  {15}},\ \bibinfo {pages} {1985--1993} (\bibinfo {year} {2000})}\BibitemShut
  {NoStop}%
\bibitem [{\citenamefont {Hetherington}\ and\ \citenamefont
  {Thorpe}(1992)}]{Hetherington1992}%
  \BibitemOpen
  \bibfield  {author} {\bibinfo {author} {\bibfnamefont {J.~H.}\ \bibnamefont
  {Hetherington}}\ and\ \bibinfo {author} {\bibfnamefont {M.~F.}\ \bibnamefont
  {Thorpe}},\ }\bibfield  {title} {\enquote {\bibinfo {title} {{The
  Conductivity of a Sheet Containing Inclusions with Sharp Corners}},}\
  }\href@noop {} {\bibfield  {journal} {\bibinfo  {journal} {Proceedings:
  Mathematical and Physical Scences}\ }\textbf {\bibinfo {volume} {438}},\
  \bibinfo {pages} {591--604} (\bibinfo {year} {1992})}\BibitemShut {NoStop}%
\bibitem [{\citenamefont {Eischen}\ and\ \citenamefont
  {Torquato}(1993)}]{Eischen1993}%
  \BibitemOpen
  \bibfield  {author} {\bibinfo {author} {\bibfnamefont {J.}~\bibnamefont
  {Eischen}}\ and\ \bibinfo {author} {\bibfnamefont {S.}~\bibnamefont
  {Torquato}},\ }\bibfield  {title} {\enquote {\bibinfo {title} {Determining
  elastic behavior of composites by the boundary element method},}\ }\href@noop
  {} {\bibfield  {journal} {\bibinfo  {journal} {Journal of applied physics}\
  }\textbf {\bibinfo {volume} {74}},\ \bibinfo {pages} {159--170} (\bibinfo
  {year} {1993})}\BibitemShut {NoStop}%
\end{thebibliography}%

\end{document}